%% file: main.tex
\documentclass{SciPost}

\input{preamble}

\begin{document}

\pagestyle{SPstyle}

\input{0-frontmatter}

\input{1-intro}
\input{2-overview}
\input{3-examples}
\input{4-conclusion}

\section*{Acknowledgements}

We gratefully acknowledge that this package, including its name, originated from an early-stage code repository developed by Maarten Van Damme.
We thank Olivier Gauthé for insightful benchmarking efforts, suggestions, and solutions, and particularly for contributions that improved the performance of CTMRG contractions and the simple update implementation.
We also acknowledge inspiring discussions with Nick Bultinck, Sander De Meyer, Anna Francuz, Zheng-Cheng Gu, Juraj Hasik, Yuchi He, Katharine Hyatt, Marek M. Rams, Norbert Schuch, Wei Tang, Laurens Vanderstraeten, Bram Vanhecke, Daan Verraes, Frank Verstraete, Xingyu Zhang, and Yintai Zhang.

\paragraph{Author contributions}

Over the course of its development, PEPSKit.jl has benefited from many discussions, suggestions, bug reports, and contributions from the broader tensor network and Julia communities spread over a variety of universities, countries, and cultures.
The authors would like to acknowledge the collaborative and open nature of the ecosystem surrounding tensor network software, which has significantly shaped the design and implementation of the package.

P.B. implemented the variational optimization and fixed-point differentiation framework, gauge-fixing procedures, and different flavors of the CTMRG algorithms.
He wrote \cref{sec:j1j2} and performed the optimization benchmark therein, and prepared the figures of the manuscript.
L.B. contributed to the stability and performance of the variational optimization, in particular improving the gauge-fixing and fixed-point differentiation methods, implemented support for boundary MPS contractions and added examples of applications to (3D) statistical mechanics.
He wrote \cref{sec:ipeps,sec:contraction,sec:optimization,sec:triangular_hubbard,sec:benchmarks}, performed the benchmarks reported in the latter two sections, and prepared all diagrams.
Z.Y. implemented the Trotter-based time evolution module, including support for finite-temperature simulations, and wrote \cref{sec:time-evolution,sec:finite-T-spin-model} of the manuscript.
G.F. developed infrastructure for large-scale variational iPEPS simulations, with a particular focus on handling internal symmetries, that was then used in the examples and benchmarks of this manuscript.
He also provided extensive testing and user feedback during development.
J.H. provided much of the theoretical framework underlying the methods implemented in PEPSKit.jl, as well as the supporting software infrastructure for (symmetric) tensor contraction primitives, optimization algorithms, and (automatic) differentiation rules for factorization routines.
L.D. initiated and coordinated the project, designed and implemented the core framework of PEPSKit.jl.
He oversaw code review throughout the project, organized technical discussions, and wrote \cref{sec:intro,sec:conclusion} of the manuscript.
All authors contributed to discussions, manuscript revisions, and approved the final version of the manuscript.

\paragraph{Funding information}
P.B.\ is supported by the European Union’s Horizon 2020 research and innovation programme through Grant No. 863476 (ERC-CoG SEQUAM).
L.B. and J.H.\ are supported by the European Union’s Horizon 2020 research and innovation programme through Grant No. 101125822 (ERC-CoG GaMaTeN).
Z.Y.\ is supported by funding from Hong Kong’s Research Grants Council (CRF C7015-24G, CRS HKU701/24).
G.F.\ is supported by a doctoral fellowship from the Research Foundation – Flanders (FWO) (Grant No. 11A6W25N) and acknowledges EuroHPC access to MareNostrum5 as BSC, Spain (EHPC-DEV-2025D12-057).
Parts of the computational results have been achieved using the Austrian Scientific Computing (ASC) infrastructure.
L.D.\ is supported by the Flatiron Institute.
The Flatiron Institute is a division of the Simons Foundation.

\bibliography{PEPSKit.bib}

\end{document}

%% file: preamble.tex
\hypersetup{
    colorlinks,
    linkcolor={red!50!black},
    citecolor={blue!50!black},
    urlcolor={blue!80!black}
}

\usepackage[bitstream-charter]{mathdesign}
\usepackage{amsmath}
\usepackage{braket}
\usepackage{xparse}
\usepackage{graphicx}
\usepackage{subcaption}
\usepackage[export]{adjustbox}
\usepackage[capitalize]{cleveref}
\usepackage{verbatim}
\usepackage{dsfont}
\usepackage[english]{babel}
\usepackage[autostyle, english = american]{csquotes}
\MakeOuterQuote{"}
\usepackage{textcomp}
\usepackage{tikz}

\DeclareSymbolFont{usualmathcal}{OMS}{cmsy}{m}{n}
\DeclareSymbolFontAlphabet{\mathcal}{usualmathcal}

\fancypagestyle{SPstyle}{
\fancyhf{}
\lhead{\colorbox{scipostblue}{\bf \color{white} ~SciPost Physics Codebases }}
\rhead{{\bf \color{scipostdeepblue} ~Submission }}

\fancyfoot[C]{\textbf{\thepage}}
}

\newcommand{\U}[1]{\ensuremath{\mathrm{U}(#1)}}
\newcommand{\SU}[1]{\ensuremath{\mathrm{SU}(#1)}}

\def \Z{\mathbb{Z}}

\def \tr{\operatorname{tr}}

\newcommand{\cT}{\mathcal{T}}
\newcommand{\dsI}{\mathds{1}}
\DeclareMathOperator*{\argmin}{argmin}
\newcommand{\td}{\text{d}}

\NewDocumentCommand{\diagram}{ O{1.0} O{1} O{./fig/diagrams} m }{
  \vcenter{\hbox{\includegraphics[scale=#1,page=#2]{#3/#4.pdf}}}
}

%% file: 0-frontmatter.tex
\begin{center}{\Large \textbf{\color{scipostdeepblue}{
PEPSKit.jl: A Julia package for \\ projected entangled-pair state simulations\\
}}}\end{center}

\begin{center}\textbf{
Paul Brehmer\textsuperscript{1$\star$},
Lander Burgelman\textsuperscript{2$\dagger$},
Zheng-Yuan Yue\textsuperscript{3$\ddagger$},\\
Gleb Fedorovich\textsuperscript{2},
Jutho Haegeman\textsuperscript{2} and
Lukas Devos\textsuperscript{4$\S$}
}\end{center}

\begin{center}
{\bf 1} Department of Physics, University of Vienna, Boltzmanngasse 9, 1090 Vienna, Austria
\\
{\bf 2} Department of Physics and Astronomy, Ghent University,\\Krijgslaan 299, 9000 Gent, Belgium
\\
{\bf 3} Department of Physics, The Chinese University of Hong Kong,\\Sha Tin, New Territories, Hong Kong, China
\\
{\bf 4} Center for Computational Quantum Physics, Flatiron Institute,\\New York, New York 10010, USA
\\[\baselineskip]
$\star$ \href{mailto:paul.brehmer@univie.ac.at}{\small paul.brehmer@univie.ac.at}\,,\quad
$\dagger$ \href{mailto:lander.burgelman@ugent.be}{\small lander.burgelman@ugent.be}\,,\\
$\ddagger$ \href{mailto:zhengyuanyue@cuhk.edu.hk}{\small zhengyuanyue@cuhk.edu.hk}\,,\quad
$\S$ \href{mailto:ldevos98@gmail.com}{\small ldevos98@gmail.com} 
\end{center}

\section*{\color{scipostdeepblue}{Abstract}}
\textbf{\boldmath{%
We present PEPSKit.jl, a Julia package for simulating two-dimensional quantum many-body systems with infinite projected entangled-pair states (iPEPS).
PEPSKit.jl builds on the TensorKit.jl package for tensor computations and provides high-level algorithms for iPEPS simulations that support both Abelian and non-Abelian symmetries, as well as fermionic systems.
This work gives an overview of the main package features, which include support for ground-state, time-evolution, and finite-temperature simulations in systems with different physical symmetries and lattice geometries.
These capabilities are illustrated through various examples and technical benchmarks.
\vspace{1em}
}}

\vspace{\baselineskip}

\noindent\textcolor{white!90!black}{%
\fbox{\parbox{0.975\linewidth}{%
\textcolor{white!40!black}{\begin{tabular}{lr}%
  \begin{minipage}{0.6\textwidth}%
    {\small Copyright attribution to authors. \newline
    This work is a submission to SciPost Physics Codebases. \newline
    License information to appear upon publication. \newline
    Publication information to appear upon publication.}
  \end{minipage} & \begin{minipage}{0.4\textwidth}
    {\small Received Date \newline Accepted Date \newline Published Date}%
  \end{minipage}
\end{tabular}}
}}
}


\vspace{10pt}
\noindent\rule{\textwidth}{1pt}
\tableofcontents
\noindent\rule{\textwidth}{1pt}
\vspace{10pt}

%% file: 1-intro.tex
\section{Introduction}
\label{sec:intro}

Understanding and predicting the behavior of many interacting quantum particles remains one of the central challenges of modern physics.
Many pertinent questions, such as the mechanisms of high-temperature superconductivity or the classification of frustrated magnets and topological phases, remain difficult to tackle.
These difficulties are in large part due to the exponential growth of the Hilbert space of a quantum many-body system with the number of constituents, which causes the computational cost of brute-force simulation to scale exponentially with system size.
Tensor network states \cite{verstraete_matrix_2008,bridgeman_handwaving_2017,cirac_matrix_2021} address this challenge by exploiting the entanglement structure of physically relevant states.
Low-energy states of local gapped Hamiltonians satisfy an area law for their entanglement entropy \cite{hastings_area_2007,eisert_colloquium_2010}, which confines them to a small corner of Hilbert space that can be efficiently parametrized by tensor networks.

The power of this approach is most clearly demonstrated by matrix product states (MPS), which have become a standard tool for simulating 1D quantum systems \cite{white_density_1992,cirac_matrix_2021}.
MPS-based methods tend to be mature, robust, and supported by a rich ecosystem of open-source software implementations.
This enables their straightforward application to a wide range of problems.
Extending these successes to higher dimensions is, however, far from straightforward.
While MPS can in principle be applied to quasi-2D geometries such as strips or cylinders, this requires computational resources that grow exponentially with the width of the system.
This makes accurate simulations of genuinely 2D systems using MPS prohibitively expensive.

A natural generalization to 2D is provided by projected entangled-pair states (PEPS) \cite{verstraete_valencebond_2004}, and in particular their infinite-system variant (iPEPS).
This class of tensor network states directly encodes translationally invariant 2D quantum states in the thermodynamic limit.
iPEPS methods have been successfully applied to a range of spin and fermionic lattice models, including ground-state and excited-state optimization \cite{vanderstraeten_simulating_2019,ponsioen_excitations_2020,ponsioen_automatic_2022,ponsioen_improved_2023}, real- and imaginary-time dynamics \cite{corboz_competing_2014,czarnik_time_2019,espinoza_spectral_2024}, and finite-temperature calculations \cite{sinha_finite_2022,zhang_finite_2026}.

Despite the impressive results, iPEPS remains far less adopted than MPS in practice.
This stems from two compounding factors: the inherent theoretical complexity of 2D tensor networks and the resulting difficulty in developing stable, efficient software implementations.
The theoretical complexity is primarily caused by loops in 2D tensor networks.
Exact contraction of such tensor networks is \#P-hard in general \cite{schuch_computational_2007}, necessitating approximate methods like the corner transfer matrix renormalization group (CTMRG) \cite{nishino_corner_1996,orus_simulation_2009} or boundary-MPS methods \cite{verstraete_renormalization_2004,jordan_classical_2008,vanderstraeten_variational_2022}.
Furthermore, the loops preclude the existence of exact canonical forms \cite{vidal_efficient_2003,schollwock_density_2011}, which are essential for both the numerical stability and efficiency enjoyed by 1D algorithms.
These theoretical limits in turn create a steep practical barrier.
Typical iPEPS algorithms suffer from an inherently higher computational complexity, with cost scalings reaching up to $D^{12}$ in contrast to $D^3$ in 1D.
Here, the so-called bond dimension $D$ is a parameter that controls the approximation error of the methods.
Additionally, the reliance on approximate contractions introduces numerical instabilities that are often not present in 1D algorithms.
Consequently, it is far more difficult to achieve a stable implementation of iPEPS algorithms than their 1D counterparts.

Moreover, as a younger and more active field, better solutions to these challenges in iPEPS algorithms are being developed at a rapid pace.
Therefore, a state-of-the-art toolbox necessarily draws from a rapidly expanding collection of both high-level improvements such as preconditioners \cite{zhang_accelerating_2026} and gauging algorithms \cite{tang_gauging_2025}, medium-level core components such as Krylov solvers \cite{gutknecht_brief_2007}, randomized linear algebra \cite{murray_randomized_2023}, and automatic differentiation (AD) \cite{liao_differentiable_2019} with support for fixed-point differentiation \cite{christianson_reverse_1994,liao_differentiable_2019}, as well as low-level dedicated kernels for a variety of HPC hardware.
Building, maintaining, and extending a toolbox for iPEPS algorithms therefore requires time, effort, and expertise in fields that extend well beyond many-body physics.
Consequently, the open-source software landscape for iPEPS is considerably less developed than that for MPS, raising the barrier to entry for new researchers.
This inevitably slows the development of the field.
That software barrier has begun to attract direct attention, with several open-source packages emerging in recent years, including Ace-TN, ITensor, peps-torch, quimb, TeNeS, variPEPS, and YASTN~\cite{ace_richards_2025, fishman_itensor_2022, hasik_pepstorch_2024, gray_quimb_2018, motoyama_tenes_2022,motoyama_tenes_2025, naumann_introduction_2024, rams_yastn_2025}.

In this work, we present PEPSKit.jl, an open-source Julia \cite{bezanson_julia_2017} package for iPEPS algorithms designed to lower software barriers and address theoretical challenges in 2D tensor network simulations.
PEPSKit.jl is part of the broader QuantumKitHub ecosystem \cite{quantumkithub_2024}, which provides shared infrastructure for symmetric tensor algebra, linear algebra routines, and Krylov-based solvers.
This foundation allows PEPSKit.jl to provide a unified, optimized framework for ground-state optimization, time evolution, and finite-temperature calculations.

A distinguishing feature of PEPSKit.jl is its comprehensive support for Abelian and non-Abelian internal symmetries, as well as fermionic tensors \cite{mortier_fermionic_2025} via TensorKit.jl \cite{devos_tensorkitjl_2025}. 
Notably, this framework avoids having to manually resolve leg crossings with swap gates for fermionic systems \cite{corboz_fermionic_2009}, and exploiting these symmetries can yield significant computational speedups (see \Cref{sec:benchmarks}). 
Additionally, for ground state optimization, PEPSKit.jl relies on automatic differentiation (AD) \cite{liao_differentiable_2019} rather than analytically derived gradients \cite{vanderstraeten_gradient_2016,corboz_variational_2016}.
By utilizing Julia's full-language AD ecosystem \cite{innes_dont_2019}, backpropagation through iterative solvers can be smoothly replaced with implicit differentiation, ensuring numerical stability and efficiency.
This approach reflects the current state of the art and drastically simplifies the integration of new models and ans\"{a}tze.
The current release targets finite-range Hamiltonians on square 2D lattices, with extensions to more general geometries planned for future work.

In joining a growing landscape of open-source PEPS software, PEPSKit.jl stands out by combining Trotter-based time evolution, AD-based optimization, and support for symmetric tensors in a single package.
Consequently, it delivers both production-ready algorithms for complex physics calculations and composable building blocks that minimize the effort required to test new algorithmic ideas.
The package is fully open-source, and the code, benchmarks, documentation as well as an extended set of runnable examples are hosted on GitHub \cite{pepskit}.

The remainder of this paper is organized as follows.
\Cref{sec:overview} provides an overview of PEPSKit.jl and its main components, situating them within the general iPEPS workflow.
\Cref{sec:examples} presents applications to several quantum lattice models with different symmetries and geometries, accompanied by benchmarks illustrating performance scaling with non-Abelian symmetry complexity.
\Cref{sec:conclusion} offers concluding remarks and an outlook on future developments.

%% file: 2-overview.tex
\section{Overview of package features}
\label{sec:overview}

PEPSKit.jl focuses on working with two particular classes of two-dimensional tensor networks: infinite projected entangled-pair states (iPEPS) and infinite projected entangled-pair operators (iPEPO).
The former provide an efficient representation of quantum states, directly in the thermodynamic limit \cite{verstraete_valencebond_2004} and exhibit an area-law scaling of their entanglement entropy by construction \cite{verstraete_criticality_2006}, while the latter can similarly represent density operators or transfer matrices of 3D partition functions. 
The functionality of PEPSKit.jl is centered around the manipulation of these objects, and can be divided into three main categories: \emph{contraction}, \emph{time evolution}, and \emph{variational optimization}.
We summarize the core structure of the library in \cref{fig:library_overview}.

\begin{figure}
    \centering
    \input{fig/library_overview.tikz}
    \caption{Overview of the software package structure. PEPSKit.jl is built using tensor objects defined in the TensorKit.jl library which generically support internal symmetries including fermionic statistics. The underlying tensor decompositions and linear algebra routines are provided through MatrixAlgebraKit.jl~\cite{matrixalgebrakit} and enjoy backwards-mode AD compatibility. Various auxiliary packages aid the implementation of specific routines such as boundary MPS methods, which draw from the MPSKit.jl~\cite{devos_mpskit_2026} library, or the optimization and differentiation functionalities which employ the Zygote.jl AD engine~\cite{innes_dont_2019}. The core data objects implemented in PEPSKit.jl can be categorized into states, operators and environments of states. Based on these objects, we provide a wide range of algorithms for contraction, time evolution and AD-based optimization.}
    \label{fig:library_overview}
\end{figure}

\subsection{The iPEPS and iPEPO ansatz}
\label{sec:ipeps}

The variational parameters defining an iPEPS or iPEPO correspond to a local tensor, which can be interpreted as a multilinear map from four virtual vector spaces to a physical vector space.
We denote this local tensor by $A$, and depict it as
\begin{align}
    \label{eq:pepso_tensor}
    A &: V_N \otimes V_E \otimes V_S^* \otimes V_W^* \to P,
    &
    A &: V_N \otimes V_E \otimes V_S^* \otimes V_W^* \to P \otimes P^*, \\
    & \diagram[1.0]{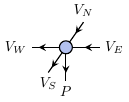}
    &
    & \diagram[1.0]{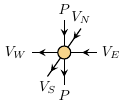}
    \nonumber
\end{align}
for the case of a PEPS and PEPO tensor, respectively.
Here, we use incoming and outgoing arrows only as a graphical tool to indicate whether the corresponding vector spaces are dual or not.
The dimension of the physical vector space $P$ is referred to as the \emph{physical dimension}, denoted as $d$.
The dimension of the virtual vector spaces $V_N$, $V_E$, $V_S$ and $V_W$ is called the \emph{bond dimension}, and we will denote it as $D$.

Given a spin system on a square lattice with a local Hilbert space $P$ at each site, we can parametrize a many-body quantum state or operator in terms of the local tensor $A$ by placing a copy of the tensor at each lattice site and contracting the virtual indices according to the lattice connectivity.
This gives rise to an iPEPS $\ket{\psi(A)}$ or iPEPO $\rho(A)$, defined as
\begin{align}
    \label{eq:pepso}
    \ket{\psi(A)} = \diagram[0.8]{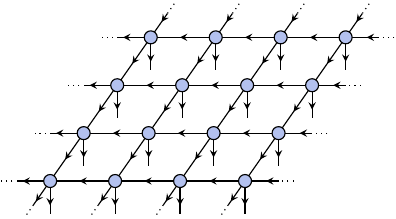},
    \quad
     \rho(A) = \diagram[0.8]{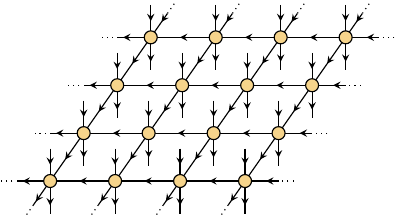}.
\end{align}
These network definitions require a consistent choice for the duality of the spaces in the domain and codomain of the local tensor map $A$ of \cref{eq:pepso_tensor}.
This is reflected in the consistency of the arrows in the network diagrams \cref{eq:pepso}, where different choices would lead to different consistent arrow configurations.
While the distinction between different choices can become important when considering tensors with internal symmetries \cite{devos_tensorkitjl_2025}, it is not relevant to this overview.
Therefore, we will drop the arrows from the diagrams from here on out to keep the notation uncluttered.

Armed with these basic definitions, we can express relevant quantities of interest as contractions of tensor networks.
For example, computing the expectation value of a local observable $O$ boils down to evaluating networks of the form
\begin{align}
    \label{eq:expval}
    \braket{\psi(A)|O|\psi(A)} = \diagram[0.75]{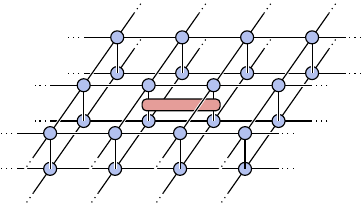}, \;
    \tr(\rho(A) O) = \diagram[0.75]{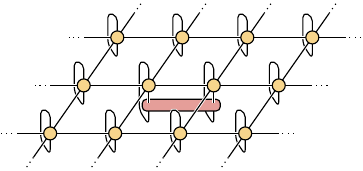}.
\end{align}
Crucially, these quantities cannot be evaluated exactly in general, but rather have to be computed by approximating the infinite parts of the network in a suitable way.
This is achieved through the use of \emph{approximate contraction algorithms}, which form the basis of all iPEPS algorithms.

Before discussing the issue of contraction in more detail, we first highlight some nuances of the simple network definitions given above.
First, instead of using a single local tensor, one can also use a unit cell of tensors as variational degrees of freedom in order to represent states with larger unit cells.
Second, the approach is not limited to square lattices, but can be applied to other regular lattices.
This can be done either by directly adapting the number of virtual spaces of the local tensors to the coordination number of the lattice, or by blocking several lattice sites together to achieve an effective square geometry.
At the moment, PEPSKit.jl only supports the latter approach, with support for the former planned for the future.
Finally, for systems with global physical symmetries, the spaces in \cref{eq:pepso_tensor} naturally correspond to graded vector spaces corresponding to a direct sum of irreducible spaces labeled by the \emph{sectors} or \emph{charges} of the symmetry.
For example, for group-like symmetries the charge sectors correspond to the irreps of the group, while for fermionic systems the charge sectors correspond to the fermion parity \cite{mortier_fermionic_2025}.
This internal structure can be exploited to significantly reduce the computational cost of iPEPS algorithms.
PEPSKit.jl provides full support for both Abelian and non-Abelian symmetries through the TensorKit.jl \cite{devos_tensorkitjl_2025} package.
We will not consider the details of these nuances for the remainder of this overview.
They will be further elaborated on later in the examples of \cref{sec:examples}.

\subsection{Contraction of 2D tensor networks}
\label{sec:contraction}

To evaluate quantities of interest, such as the expectation values in \cref{eq:expval}, we need a way to approximate the infinite parts of the network.
This problem essentially corresponds to the contraction of a classical partition function, which can be done by replacing the infinite regions by a \emph{contraction environment}.
For example, to compute an iPEPS expectation value we can approximate the corresponding norm network $\braket{\psi|\psi}$ as
\begin{equation}
    \label{eq:norm_contraction}
    \diagram[1.0]{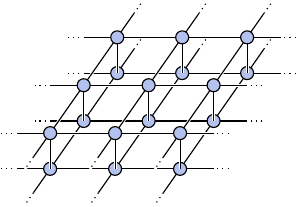}
    \to
    \diagram[1.0]{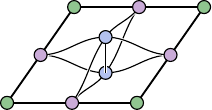}
\end{equation}
Here, the contraction environment is represented by the purple and green tensors.
In particular, we refer to the dimension of the virtual space connecting the environment tensors as the \emph{environment bond dimension}, denoted as $\chi$, which controls the accuracy of the contraction approximation.
To illustrate the different approaches to computing a contraction environment supported by PEPSKit.jl, we will focus on the example of an iPEPS norm network.
However, the same routines are directly applicable to the contraction of different effective partition functions, which can arise from classical statistical mechanics, such as $\braket{\psi|\rho|\psi}$ or $\tr(\rho^n)$.

A first class of contraction algorithms supported by PEPSKit.jl is based on the corner transfer matrix renormalization group (CTMRG) \cite{nishino_corner_1996,nishino_corner_1997,orus_simulation_2009}.
In this approach, the contraction environment consists of a set of corner and edge tensors, which are iteratively updated by absorbing layers of the network and truncating the resulting tensors in a renormalization procedure,
\begin{equation}
    \label{eq:ctmrg}
    \diagram[1.0]{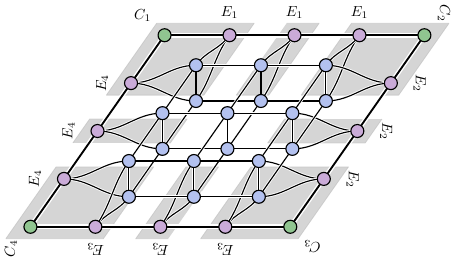}
    \to
    \diagram[1.0]{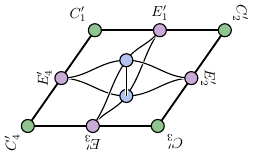}.
\end{equation}
This renormalization step is then repeated until convergence.
For bosonic systems with a spatial $C_{4v}$ point-group symmetry, which requires a single-site unit cell where the local tensor satisfies
\begin{equation}
    \label{eq:peps_tensor_symmetries}
    \diagram[1.0]{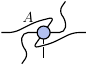} = \diagram[1.0]{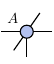},
    \quad
    \diagram[1.0]{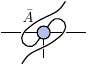} = \diagram[1.0]{peps_tensor_rhs},
\end{equation}
PEPSKit.jl implements a $C_{4v}$ CTMRG algorithm based on either a truncated Hermitian eigenvalue decomposition \cite{nishino_corner_1996} or a QR decomposition \cite{zhang_accelerating_2025}.
For systems without spatial symmetries, PEPSKit.jl implements several flavors of the generic CTMRG algorithm \cite{orus_simulation_2009,corboz_simulation_2010,corboz_competing_2014} which can handle arbitrary unit cells.

A second class of contraction algorithms available in PEPSKit.jl is the boundary MPS~\cite{haegeman_diagonalizing_2017,zauner_variational_2018} algorithm.
This approach is based on the observation that the left-hand side of \cref{eq:norm_contraction} can be viewed as an infinite power of a row-to-row transfer operator $\cT$.
To contract the norm network, we can then find the optimal MPS approximations to the dominant right and left eigenvectors of $\cT$,
\begin{align}
    \nonumber
    \diagram[1.0]{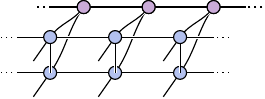}
    \propto
    \diagram[1.0]{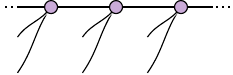},
    \\
    \diagram[1.0]{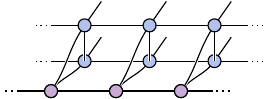}
    \propto
    \diagram[1.0]{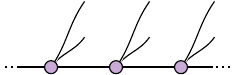}.
    \label{eq:boundarymps}
\end{align}
The remaining 1D network can then be treated using standard MPS methods, which are provided through the MPSKit.jl library \cite{devos_mpskit_2026}.
PEPSKit.jl offers access to different flavors of boundary MPS contractions, including the variational uniform matrix product state (VUMPS) algorithm \cite{zauner_variational_2018} or its power-method equivalent \cite{vanhecke_tangentspace_2021}, and a manifold-based approach using Riemannian optimization \cite{hauru_riemannian_2021} for systems with Hermitian reflection symmetry.
The boundary MPS approach also supports generic unit cells \cite{nietner_efficient_2020}.

\subsection{Time evolution based on Trotter decomposition}
\label{sec:time-evolution}

The (imaginary) time evolution of an iPEPS $\ket{\psi}$ with Hamiltonian $H$ for a period $\tau$ is given by $e^{-H\tau} \ket{\psi}$, which is calculated in PEPSKit.jl utilizing a Trotter decomposition of the evolution operator $e^{-H\tau}$.
For simplicity of demonstration, suppose $H = \sum_{R,l} H_{R,l}$ on a square lattice, where $H_{R,l}$ acts on the $l$-th nearest-neighbor (NN) bond in the unit cell at position $R$, with translation symmetry imposing $H_{R,l} = H_l$ for any $R$. 
The first-order Trotter decomposition reads
\begin{equation}
    e^{-H \tau} = (e^{-H\epsilon})^m
    \approx \biggl(
        \prod_l e^{-H_{\{l\}} \epsilon}
    \biggr)^m + O(\epsilon^2),
    \quad
    H_{\{l\}} = \sum_R H_{R,l},
\end{equation}
where $m$ is a large integer, and $\epsilon = \tau / m$. When (and only when) the unit cell size is at least $2 \times 2$, all terms in $H_{\{l\}}$ have no overlap and commute with each other, allowing a further decomposition,
\begin{equation}
    e^{-H_{\{l\}} \epsilon} = \prod_R e^{-H_{R,l} \epsilon}.
\end{equation}
Each two-site Trotter \emph{gate} $U_{R,l} = w^{-H_{R,l} \epsilon}$ acts only on the NN bond $(R,l)$.
This means we only need to know how to calculate $e^{-H_{R,l} \epsilon} \ket{\psi}$ (for real-time evolution, one simply changes $\epsilon \to i \epsilon$). Let $A$, $B$ be the two PEPS tensors connected by the bond.
After applying $U_{R,l}$, they are updated to
\begin{equation}
    U_{R,l}(A, B) = \, 
    \diagram[1.0]{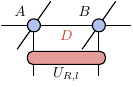}
    \, \overset{\text{SVD}}{=} \,
    \diagram[1.0]{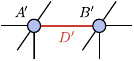}.
    \label{eq:apply-gate}
\end{equation}
However, the bond connecting $A'$ and $B'$ has a larger dimension $D' > D$.
To obtain updated $\tilde{A}, \tilde{B}$ tensors with the original bond dimension $D$, we could perform a truncated SVD in \cref{eq:apply-gate}, which minimizes the (local) difference between the patches of two tensors.
However, we aim to best approximate the (global) state instead by minimizing the cost function
\begin{equation}
    f(\tilde{A}, \tilde{B}) = \left\Vert
        \ket{\psi(A',B')} 
        - \ket{\psi(\tilde{A}, \tilde{B})}
    \right\Vert^2
    \label{eq:timeevol-cost}
\end{equation}
where $\ket{\psi(A',B')}$ and $\ket{\psi(\tilde{A},\tilde{B})}$ are obtained from the original $\ket{\psi}$ by replacing $(A, B)$ with $(A', B')$ and $(\tilde{A},\tilde{B})$, respectively.
Finally, we replace the $A, B$ tensors in all unit cells with the optimal $\tilde{A}, \tilde{B}$ to approximate $e^{-H_{\{l\}} \epsilon} \ket{\psi}$.
The update process for other NN bonds in each unit cell is performed similarly. 

\subsubsection{Bond environment and various evolution schemes}

\begin{figure}[bt]
    \centering
    \begin{subfigure}[c]{.32\textwidth}
        \centering
        \includegraphics[scale=0.85]{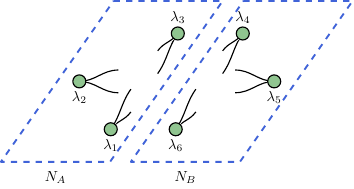}
        \caption{}
    \end{subfigure}
    \begin{subfigure}[c]{.32\textwidth}
        \centering
        \includegraphics[scale=0.9]{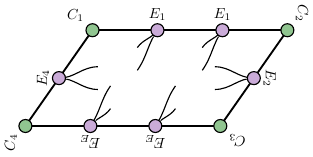}
        \caption{}
    \end{subfigure}
    \begin{subfigure}[c]{.32\textwidth}
        \centering
        \includegraphics[scale=0.9]{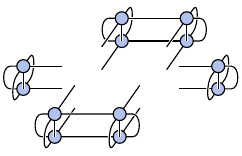}
        \caption{}
    \end{subfigure}
    \caption{Bond environment tensor for (a) the simple update scheme (which can be decomposed to two disconnected positive parts $N_A, N_B$), (b) the full update scheme, and (c) one variation of neighborhood tensor update scheme, assuming that only tensors that are nearest neighbors of $A, B$ contribute to the bond environment.}
    \label{fig:su-ntu-fu-bondenv}
\end{figure}

We define the \emph{bond environment tensor} $N_{AB}$ surrounding the bond $(A,B)$ as 
\begin{equation}
    N_{AB} = \diagram[1.0]{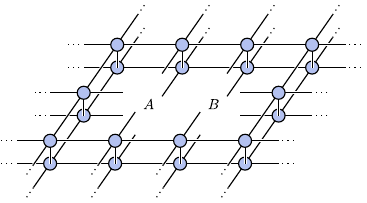},
    \label{eq:bondenv}
\end{equation}
which is obtained by contracting the entire $\braket{\psi|\psi}$ network except $A, B$ and their conjugates.
By definition, $N_{AB}$ is a positive operator.
Then the cost function \cref{eq:timeevol-cost} is equal to
\begin{equation}
    f(\tilde{A},\tilde{B})
    = (A' B' - \tilde{A} \tilde{B})^\dagger
    (N_{AB} \otimes \mathbb{1}_A \otimes \mathbb{1}_B) (A' B' - \tilde{A} \tilde{B}),
\end{equation}
where $\mathbb{1}_A, \mathbb{1}_B$ are identity operators acting on the physical space at sites $A$, $B$ respectively. 
Minimizing \cref{eq:timeevol-cost} is then equivalent to finding the weighted low-rank approximation \cite{srebro_weighted_2003} $\tilde{A} \tilde{B}$ of $A'B'$  that has rank $D < D'$, with the weight tensor $N_{AB} \otimes \mathbb{1}_A \otimes \mathbb{1}_B$.
In practice, $N_{AB}$ cannot be evaluated exactly, leading to different approximation schemes.
Currently, PEPSKit.jl implements the \emph{simple update} (SU) scheme \cite{jiang_accurate_2008}, which is a generalization of the one-dimensional time-evolving block decimation (iTEBD) algorithm \cite{vidal_classical_2007}. 
It approximates $N_{AB}$ by the tensor product of bond weights $\{\lambda\}$ (\cref{fig:su-ntu-fu-bondenv}a), each of which is a diagonal matrix with positive diagonal elements. 
In this simple case, $N_{AB}$ is a tensor product of two disconnected parts $N_A, N_B$, each being positive and contracting with only $A$ or $B$.
The problem is then reduced to finding the \emph{unweighted} low-rank approximation of $(\sqrt{N_A} \mathbb{1}_A A')(\sqrt{N_B} \mathbb{1}_B B')$.
The solution is immediately given by a truncated SVD, which simultaneously updates the bond weight between $A, B$ with the obtained singular value spectrum.

PEPSKit.jl will support more accurate approximation schemes of $N_{AB}$ in the near future.
Firstly, there is the \emph{full update} (FU) scheme \cite{phien_infinite_2015, haghshenas_u1_2018}, which approximates $N_{AB}$ with converged CTMRG tensors (\cref{fig:su-ntu-fu-bondenv}b).
Secondly, there are the different variations of the \emph{neighborhood tensor update} (NTU) scheme \cite{dziarmaga_time_2021, andrew_beyond_2025}, which assumes that only tensors in the proximity of the updated bond contribute to $N_{AB}$ (\cref{fig:su-ntu-fu-bondenv}c).
In these cases, $N_{AB}$ does not have the simple form $N_A \otimes N_B$, and one has to minimize \cref{eq:timeevol-cost} with iterative methods.

\subsubsection{Hamiltonians with next-nearest-neighbor terms}

\begin{figure}[bt]
    \centering
    \includegraphics[scale=1.0]{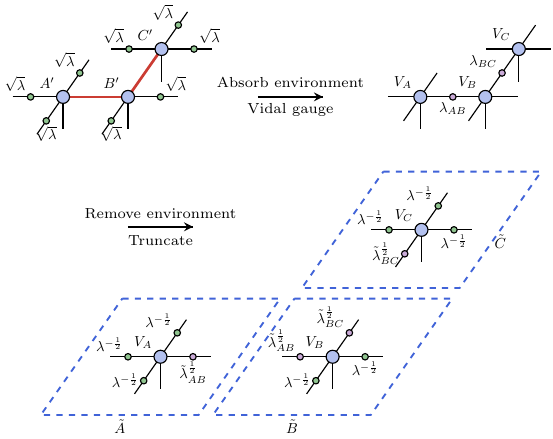}
    \caption{Truncation of internal virtual bonds (red) in the 3-site cluster regarded as an open-boundary MPS in the simple update scheme with 3-site MPO gates. $\tilde{\lambda}_{AB}, \tilde{\lambda}_{BC}$ are obtained by truncating $\lambda_{AB}, \lambda_{BC}$ respectively.}
    \label{fig:su-mpo-truncate}
\end{figure}

In general, evolution with next-nearest-neighbor (NNN) gates can be handled by a conversion to a matrix product operator (MPO) along a shortest three-site path on the lattice, which acts on the corresponding iPEPS tensors, regarded as a three-site open-boundary MPS (often referred to as a \emph{cluster}),
\begin{equation}
    \diagram[1.0]{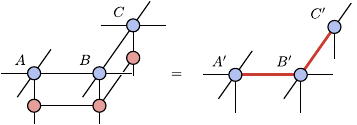}.
    \label{eq:apply-mpo}
\end{equation}
The updated internal virtual bonds in the cluster are then truncated sequentially.
For simple update, PEPSKit.jl adopts an alternative approach that truncates all updated bonds at once, which is illustrated in \cref{fig:su-mpo-truncate}.
First, square roots of the bond weights on the open virtual legs are absorbed into the cluster to account for the environment around it. 
Next, the cluster is converted to the \emph{Vidal gauge} \cite{vidal_classical_2007, tindall_gauging_2023}, followed by a truncation of the weights on the internal virtual bonds.
Finally, the iPEPS tensors are updated after removing the bond weights on all open virtual legs.
This procedure can be easily generalized to handle Trotter gates acting on arbitrary open paths along the lattice edges.
The support for NNN terms allows PEPSKit.jl to simulate models on the triangular lattice with a mapping to the square lattice, which converts some NN gates on the triangular lattice to NNN gates on the square lattice.

\subsubsection{Finite temperature}

With minor modifications, imaginary time evolution algorithms for iPEPS can be used to study quantum systems at finite temperature $\beta = 1/T$. 
At thermal equilibrium, a quantum system is in a mixed state described by the (unnormalized) density operator $\rho(\beta) = e^{-\beta H}$.
In particular, at infinite temperature $\beta = 0$, $\rho(\beta)$ reduces to the identity operator, which can be expressed as a trivial iPEPO with bond dimension $D = 1$. 
Therefore, an iPEPO representation of $\rho(\beta)$ can be obtained from the imaginary time evolution of the trivial iPEPO at $\beta = 0$, regarding it as a purified iPEPS $\ket{\rho(\beta)}$ with an extra ancilla physical leg.
The expectation value of an operator $O$ at temperature $2\beta$ can be obtained from
\begin{equation}
    \braket{O}(2\beta)
    = \frac{
        \braket{\rho(\beta)|O|\rho(\beta)}
    }{\braket{\rho(\beta)|\rho(\beta)}}.
\end{equation}
Alternatively, without purifying $\rho(\beta)$, we can obtain $\braket{O}$ at $\beta$ from
\begin{equation}
    \braket{O}(\beta)
    = \frac{\tr[\rho(\beta) O]}{\tr[\rho(\beta)]},
\end{equation}
which is significantly less costly to compute, at the cost of being less accurate and losing guaranteed positive-definiteness of the (reduced) density matrices.

\subsection{Variational optimization with automatic differentiation}
\label{sec:optimization}

Given a Hamiltonian $H$, a key use of the iPEPS ansatz is to find the best iPEPS approximation of its ground state.
If we denote the variational parameters that make up the local PEPS tensor\footnote{For systems with non-trivial unit cells made up of multiple local tensors, we can collect all of their corresponding variational parameters into a single variable $p$.} by $p$, such that $A \equiv A(p)$, this problem can be formulated as a variational optimization of the energy with respect to $p$,
\begin{equation}
    \label{eq:variational_optimization}
    p = \argmin_p E(p) = \argmin_p
    \frac{\braket{\psi(p) | H | \psi(p)}}{\braket{\psi(p)|\psi(p)}}.
\end{equation}
For a Hamiltonian defined as a sum of local terms that act on a few neighboring sites, the variational problem for the exact energy $E$ of \cref{eq:variational_optimization} can be rephrased in terms of an approximate energy density.
This consists for example of NN terms of the form
\begin{align}
    \label{eq:peps_energy_density}
    e(x, p) = \diagram[1.0]{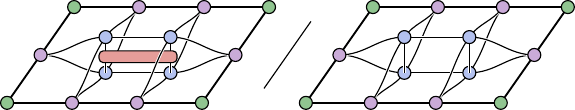},
\end{align}
but PEPSKit.jl also supports terms with a larger support, as long as the resulting network composed of the local patch surrounded by the environment tensors can be evaluated exactly with the computational resources available.
We denote the contraction environment represented by the purple and green tensors in \cref{eq:peps_energy_density} as $x$, which always implicitly depends on the variational parameters $p$ of the iPEPS, i.e., $x \equiv x(p)$.
To minimize $e(x,p)$, we can compute its gradient with respect to $p$ using the chain rule,
\begin{equation}
    \label{eq:peps_gradient_chain_rule}
    \frac{\td e}{\td p}(p) = \partial_x e(x, p) \, \partial_p x + \partial_p e(x, p),
\end{equation}
and supply it to a suitable gradient-based optimization scheme \cite{nocedal_updating_1980}.
In PEPSKit.jl we evaluate this gradient using the framework of AD, more specifically via \emph{reverse-mode AD} \cite{innes_dont_2019}, using two different methods described below.

\subsubsection{Direct differentiation through iterations}

The first approach used in PEPSKit.jl for evaluating \cref{eq:peps_gradient_chain_rule} is a direct application of reverse-mode AD.
To understand this, it is helpful to introduce \emph{adjoint variables}, which are partial derivatives of the computation output with respect to one of its variables.
In the current context, the computation output is the approximate energy density $e$.
The adjoints of the contraction environment $x$ and the variational parameters $p$ are then $\breve{x} = \partial_x e$ and $\breve{p} = \partial_p e$, respectively.
With these defnitions, \cref{eq:peps_gradient_chain_rule} can be rewritten as
\begin{equation}
    \label{eq:peps_gradient_chain_rule_with_adjoint}
    \frac{\td e}{\td p}(p)
    = \breve{x} \, \partial_p x + \breve{p}.
\end{equation}
Here, the key step is the multiplication of the Jacobian $\partial_p x$ with the adjoint vector $\breve{x}$ to its left, known as a \emph{vector-Jacobian product} (VJP).

The contraction algorithm used to compute the environment $x(p)$ is an iterative procedure where we apply an iterating function $f(x, p)$ to an initial guess $x_0$ until convergence is reached, $x^* = f(x^*, p)$.
We refer to the converged value $x^*$ as the \emph{fixed point} of the iterating function $f$.
This procedure gives rise to intermediate objects $x_i$, defined as $x_1 = f(x_0, p)$, $x_2 = f(x_1, p)$, $\dots$, $x_{n-1} = f(x_{n-2}, p)$, and finally $x^* := x_n = f(x_{n-1}, p)$.
Using the chain rule, we have
\begin{align}
    \label{eq:black_box_ad}
    \breve{x} \, \partial_p x 
    = \breve{x} \, \sum_{k=1}^n \left(\left( \prod_{m=1}^{k-1} \partial_x f(x_{n-m}, p) \right) \partial_p f(x_{n-k}, p)\right),
\end{align}
where $\partial_x f$ and $\partial_p f$ can be obtained from automatic differentiation. In other words, this approach evaluates \cref{eq:peps_gradient_chain_rule_with_adjoint} by directly propagating adjoints backwards through all $n$ iterations of $f$.
It has the advantage that as long as we ensure that an AD engine can automatically compute the basic derivatives $\partial_x e$ and $\partial_p e$, as well as the VJP actions of $\partial_x f$, $\partial_p f$, we can obtain \cref{eq:black_box_ad} from a \enquote{black-box} application of reverse-mode AD to the entire contraction algorithm.
However, the downside is that all intermediate objects $x_i$ need to be saved in memory unless a checkpointing strategy is employed~\cite{griewank_evaluating_2008} which, in turn, increases the computational cost.

\subsubsection{Fixed-point differentiation}

The second approach for evaluating \cref{eq:peps_gradient_chain_rule_with_adjoint} used in PEPSKit.jl avoids the memory overhead of backpropagating through all iterations of the contraction algorithm by leveraging an \emph{implicit} characterization of the contraction environment.
In particular, we can fully characterize the converged environment $x^*$ as the solution to the \emph{fixed-point equation}
\begin{equation}
    \label{eq:fixed_point_equation}
    x = f(x, p).
\end{equation}
By applying the implicit function theorem \cite{christianson_reverse_1994} to this equation, we can obtain the VJP action of $\partial_p x$ as
\begin{equation}
    \label{eq:fixed_point_differentiation_vjp}
    \breve{x} \, \partial_p x = \breve{x} \left(\dsI - \partial_x f\right)^{-1} \left(\partial_p f\right).
\end{equation}
Crucially, the VJP action of $\partial_x f$ and $\partial_p f$ in this expression are evaluated only at the fixed point $x^*$, which greatly reduces the memory overhead in the gradient evaluation.
This approach is therefore called \emph{fixed-point differentiation} \cite{liao_differentiable_2019,naumann_introduction_2024,francuz_stable_2025}.

To evaluate \cref{eq:fixed_point_differentiation_vjp} in practice, we can first use a standard linear solver to compute the solution $\breve{y}$ to the linear problem
\begin{equation}
    \label{eq:fixed_point_differentiation_linear_problem}
    \breve{y} \left( \partial_x f - \dsI \right) = \breve{x}.
\end{equation}
From this, we obtain the VJP action of $\partial_p f$ as
\begin{equation}
    \label{eq:fixed_point_differentiation_vjp_final}
    \breve{x} \, \partial_p x = - \breve{y} \, \partial_p f.
\end{equation}
Alternatively, we can rewrite the inverse ($\dsI - \partial_x f)^{-1}$ in \cref{eq:fixed_point_differentiation_vjp} as a geometric series,
\begin{equation}
    \label{eq:fixed_point_differentiation_vjp_series}
    \breve{x} \, \partial_p x
    = \sum_{k=0}^\infty \breve{x} \left( \partial_x f \right)^k \partial_p f.
\end{equation}
This can be evaluated using a simple power method, where we add terms until convergence is reached \cite{christianson_reverse_1994}.
PEPSKit.jl offers different implementations for both of these approaches to fixed-point differentiation.

Finally, we note that to use fixed-point differentiation, we require that applying a single iteration $f$ of the contraction algorithm at the fixed-point environment $x^*$ produces tensors that are \emph{element-wise} equal to those of $x^*$ within the convergence tolerance of the algorithm.
To ensure element-wise convergence, PEPSKit.jl implements suitable gauge-fixing routines for the contraction environment \cite{francuz_stable_2025}.

%% file: fig/library_overview.tikz
\begin{tikzpicture}[
    box/.style={
        draw,
        rounded corners=3pt,
        minimum width=3.0cm,
        minimum height=1.05cm,
        line width=0.75pt,
        font=\footnotesize,
        fill=white,
    },
    arrow/.style={
        ->,
        >=stealth,
        line width=0.75pt,
    },
    groupbox/.style={
        draw,
        dashed,
        rounded corners=8pt,
        fill opacity=0.85,
        line width=0.8pt,
    },
]

\definecolor{jlblue}{rgb}{0.251,0.388,0.847}
\definecolor{jlgreen}{rgb}{0.133,0.545,0.133}
\definecolor{jlpurple}{rgb}{0.584,0.345,0.698}
\colorlet{topcol}{jlblue}
\colorlet{middlecol}{jlpurple}
\colorlet{botcol}{jlgreen}

\node[inner sep=0pt] (logo) at (-4.9,5.4) {\includegraphics[width=2.0cm]{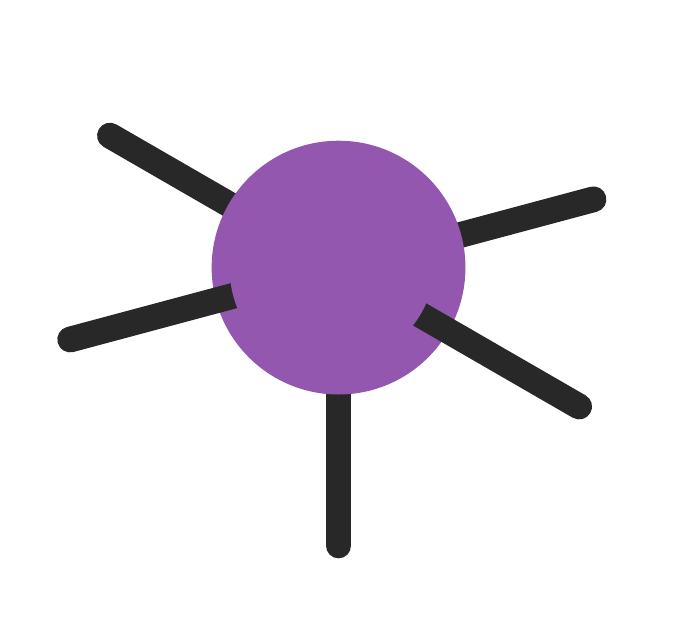}};
\node[anchor=south, font=\large] at (-4.9,3.9) {\textbf{\textsf{PEPSKit.jl}}};

\node[groupbox, draw=topcol!60, fill=topcol!30, minimum width=9.8cm, minimum height=2.2cm] (groupTop) at (2, 5) {};
\node[text=topcol, rotate=90, anchor=south] at (6.85,5) {imports};
\node[text=topcol!50, rotate=90, font=\large, yscale=2.0, xscale=1.5] at (0,3.5) {$\Leftarrow$};

\node[box, align=center] (A) at (0, 5) {
    \textbf{Tensors \& symmetries}\\\textbf{with AD compatibility:}\\
    \texttt{MatrixAlgebraKit.jl},\\\texttt{TensorKit.jl}
};
\node[box, align=center] (A) at (4, 5) {
    \textbf{Auxiliary:}\\
    \texttt{MPSKit.jl}, \texttt{Zygote.jl}\\
    \texttt{KrylovKit.jl}, \dots
};

\node[groupbox, draw=middlecol!60, fill=middlecol!30, minimum width=13.8cm, minimum height=2.2cm] (groupMiddle) at (0, 2) {};
\node[text=middlecol, rotate=90, anchor=south] at (6.85,2) {objects};
\node[text=middlecol!50, rotate=90, font=\large, yscale=2.0, xscale=1.5] at (0,0.5) {$\Leftarrow$};

\node[box, align=center] (B1) at (-4, 2) {
    \textbf{States:}\\
    infinite PEPS,\\
    partition functions, \dots
};
\node[box, align=center] (B2) at (0, 2) {
    \textbf{Operators:}\\
    local observables, PEPOs, \dots
};
\node[box, align=center] (B3) at (4, 2) {
    \textbf{Environments:}\\
    CTMRG envs., \dots
};

\node[groupbox, draw=botcol!60, fill=botcol!30, minimum width=13.8cm, minimum height=2.2cm] (groupBottom) at (0,-1) {};
\node[text=botcol, rotate=90, anchor=south] at (6.85,-1) {algorithms};

\node[box, align=center] (C1) at (-4,-1) {
    \textbf{Contraction:}\\
    CTMRG variants,\\
    boundary MPS variants,\\
    belief propagation, \dots
};
\node[box, align=center] (C3) at (0,-1) {
    \textbf{Time evolution:}\\
    Simple Update, \dots
};
\node[box, align=center] (C2) at (4,-1) {
    \textbf{AD-based optimization:}\\
    implicit differentiation,\\
    naive backpropagation, \dots
};

\end{tikzpicture}

%% file: 3-examples.tex
\section{Examples}
\label{sec:examples}

To illustrate the methods outlined in \cref{sec:overview}, we present a collection of examples that demonstrate the capabilities of PEPSKit.jl in this section.
We focus on applications of imaginary time evolution and variational optimization algorithms to spin and fermionic lattice models, where we characterize both the accuracy and performance of the package.

\subsection{Finite-temperature simulation of spin models}
\label{sec:finite-T-spin-model}

\begin{figure}
    \centering
    \includegraphics[width=0.95\linewidth]{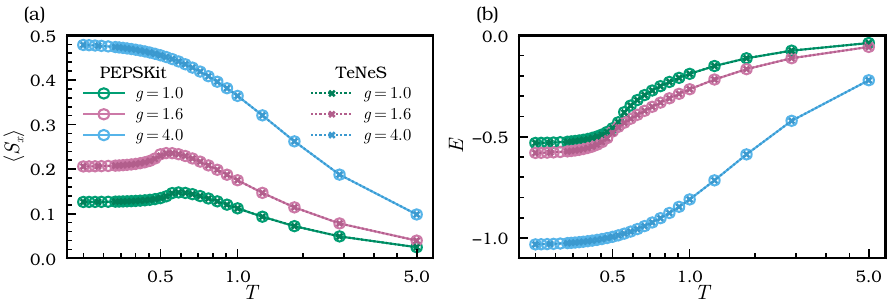}
    \caption{Finite temperature $x$-magnetization $\braket{S_x}$ and energy per site $E$ of the transverse field Ising model from Eq. \eqref{eq:tfising} with $J = 1/4$ obtained using simple update with the same settings as the TeNeS simulations of Ref.~\citenum{motoyama_tenes_2025}.
    The Trotter evolution step is $\Delta\beta = 0.02$ for the first 50 steps, $0.01$ for the following 200 steps, and $0.1$ for the final 10 steps.
    The bond dimensions are chosen in accordance with~\cite{motoyama_tenes_2025} at $D = 10$ for the iPEPO and $\chi = 16$ for the CTMRG environment.}
    \label{fig:tfising-finiteT}
\end{figure}

We begin by demonstrating the finite-temperature simulation functionality with spin models on the square lattice.
The first example is the transverse field Ising model, 
\begin{equation}
    H = -J \biggl(
        \sum_{\braket{i,j}} Z_i Z_j + g \sum_i X_i
    \biggr),
    \label{eq:tfising}
\end{equation}
where $X, Z$ are Pauli matrices.
We calculate the energy per site and the transverse magnetization for various values of the transverse field $g$, and see excellent agreement with results obtained from simple update implemented in the TeNeS package \cite{motoyama_tenes_2025} (Fig. \ref{fig:tfising-finiteT}). 

\begin{figure}
    \centering
    \includegraphics[width=0.95\linewidth]{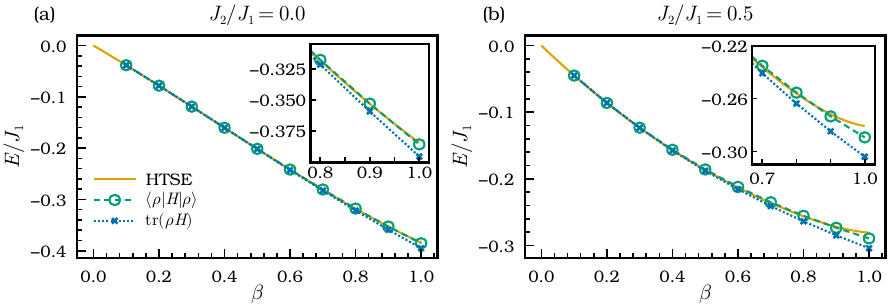}
    \caption{Energy of the $J_1$-$J_2$ model at finite temperature for (a) $J_2/J_1 = 0$ and (b) $J_2/J_1 = 0.5$, obtained from simple update with \SU{2} symmetry. The Trotter evolution step size is chosen at $\Delta\beta = 0.001$. The bond dimensions are $D \approx 7$ for the iPEPO, and $\chi = 21$ for the CTMRG environment. HTSE labels the energy obtained from the high-temperature series expansion $E = \sum_n \beta^n \sum_m e_{mn} (J_2/J_1)^m$, where the coefficients $e_{mn}$ are given in Ref.~\citenum{rosner_high_2003} up to order $n = 9$.}
    \label{fig:j1j2-finiteT}
\end{figure}

The second example is the spin-$\frac{1}{2}$ $J_1$-$J_2$ model \cite{dagotto_phase_1989}, with Hamiltonian
\begin{equation}\label{eq:j1j2}
    H = J_1 \sum_{\braket{i,j}} \vec{S}_i \cdot \vec{S}_j 
    + J_2 \sum_{\braket{\!\braket{i,j}\!}} \vec{S}_i \cdot \vec{S}_j,
\end{equation}
where $\braket{i,j}$ and $\braket{\!\braket{i,j}\!}$ sum over NN and NNN bonds in the square lattice, and the $\vec{S}_i = (X, Y, Z)$ are three-component vectors of Pauli matrices. 
At high temperatures, the spontaneous breaking of the spin rotation symmetry is suppressed by thermal fluctuations, which allows us to impose the non-Abelian \SU{2} symmetry to speed up calculations.
Fig. \ref{fig:j1j2-finiteT} shows the energy per site for $J_2 / J_1 = 0$ and $0.5$ calculated from simple update, which matches the result of the high-temperature series expansion \cite{rosner_high_2003}.

\subsection{Variational optimization for the square lattice \texorpdfstring{$J_1$-$J_2$}{J1-J2} model}
\label{sec:j1j2}

\begin{figure}
    \centering
    \includegraphics[width=0.95\linewidth]{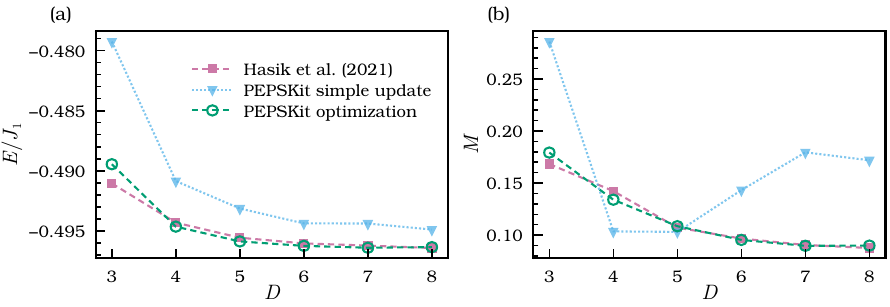}
    \caption{Energies (a) and magnetizations (b) of the square lattice $J_1$-$J_2$ model at $J_2/J_1 = 0.5$.
    We compare the measurements obtained from gradient-based optimizations with simple update runs as well as optimization data of Hasik et al.~\cite{hasik_investigation_2021}.
    The iPEPS optimizations were run at fixed CTMRG environment dimension after which observables were evaluated at an increased $\chi$.
    The precise environment dimensions for the observable evaluations were chosen in accordance with~\cite{hasik_investigation_2021} at $\chi = (144, 192, 200, 252, 245, 270)$ for the respective bond dimensions $D=3,\dots,8$.}
    \label{fig:j1j2-energy-magn-optimization}
\end{figure}

As the next example, we again treat the spin-$\frac{1}{2}$ square lattice $J_1$-$J_2$ model from \cref{eq:j1j2} in the highly-frustrated regime $J_2 / J_1 = 0.5$, but now at zero temperature.
To speed up calculations, we exploit the internal \U{1} spin-rotation symmetry of the Hamiltonian and hence choose a \U{1}-symmetric ansatz for the iPEPS and CTMRG environment tensors.
To determine the \U{1} charges on the virtual spaces, we begin by using simple update to evolve a random iPEPS with unit cell size $2 \times 2$ and an initial $J_2 = 0$, which is gradually increased to $0.5$ during the evolution.
States with larger $D$ are obtained by evolving small $D$ states with a relaxed SVD truncation scheme, hence retaining more singular values.
The \U{1} charges are dynamically selected during the SVD truncation process.
This way of selecting the relevant charges is computationally cheap and readily available, in particular in comparison with more involved ways of extracting \U{1} charges~\cite{hasik_investigation_2021}, and yields comparable optimization results.

The simple update states serve as initializations for variational optimization supported by AD-based energy gradients, where the virtual charge distributions of the evolved iPEPS are kept fixed.
The environment tensors are initialized by contracting the evolved iPEPS using a CTMRG algorithm with a fixed SVD truncation rank $\chi$, starting from a low-dimensional random environment where the virtual spaces populate the $\{0, -1, 1\}$ \U{1} sectors.
In this way, the appropriate charge distributions for the environment virtual spaces are automatically selected during CTMRG convergence.
After this initialization, the environment virtual spaces are then kept fixed throughout the optimization.
In order to avoid local minima at the initial optimization stage, we perturb the evolved iPEPS with Gaussian noise, where the noise amplitude is chosen in the order of the median of all PEPS tensor elements.
The gradient-based optimization is performed using the L-BFGS algorithm~\cite{nocedal_updating_1980,liu_limited_1989} with Hager-Zhang line search~\cite{hager_new_2005}.
We note that fairly high environment dimensions $\chi \sim nD^2$ with substantial prefactors $n$ are required during optimization to ensure that the line search algorithm works reliably.
When successive energy values are sufficiently converged --- at least below $\Delta E < 10^{-7}$, depending on the bond dimension --- we re-gauge the iPEPS by performing trivial simple update steps~\cite{lubasch_algorithms_2014} to rule out that the variational energy was undershot due to the optimizer artificially exploiting virtual gauge degrees of freedom~\cite{tang_gauging_2025}.
Note that we here resort to the absolute energy difference as a convergence measure since the energy gradient norm will remain moderate ($\|\nabla E\| > 10^{-4}$) throughout the optimization, especially at higher iPEPS bond dimensions, even when observables converge to high accuracies.
While full variational optimization to small gradient norms could be achieved by (sometimes greatly) increasing the environment bond dimension for a given iPEPS bond dimension, the straightforward convergence of the energy was sufficient for this simple benchmark.

We perform optimizations across a range of bond dimensions $D=3,\dots,8$ and measure the energy and magnetization $M = |\langle \vec{S} \rangle| = |\langle S_z \rangle|$, where the last equality follows since we use manifestly \U{1}-symmetric tensors.
In order to perform robust measurements, we evaluate the observables using CTMRG environments converged with an increased environment dimension after the optimization.
As shown in \cref{fig:j1j2-energy-magn-optimization}, the obtained energy and magnetization values are in good agreement with the previous large-scale iPEPS study in Ref.~\cite{hasik_investigation_2021}.

\subsection{The Fermi-Hubbard model on the triangular lattice}
\label{sec:triangular_hubbard}

We can extend our applications beyond spin models, and use PEPSKit.jl to perform variational optimization of the ground state of the Fermi-Hubbard model on the triangular lattice. The corresponding Hamiltonian is
\begin{equation}
    H = -t \sum_{\braket{i,j},\sigma} (
        \hat{c}^\dagger_{i\sigma} \hat{c}_{j\sigma} + h.c. 
    ) + U \sum_i \sum_{\sigma < \tau} \hat{n}_{i\sigma} \hat{n}_{i\tau}
    - \mu \sum_i \hat{n}_i,
    \label{eq:fermi-hubbard}
\end{equation}
where $\hat{c}_{i\sigma}$ is the fermionic annihilation operator for spin $\sigma \in \{1, 2\}$ (representing spin-up and spin-down) at site $i$, $\hat{n}_{i\sigma} = \hat{c}^\dagger_{i\sigma} \hat{c}_{i\sigma}$ is the corresponding number operator, and $\hat{n}_i = \hat{n}_{i,1} + \hat{n}_{i,2}$.

To simulate this model using the tools outlined in \cref{sec:overview}, we rewrite the NN interactions on the triangular lattice as NNN interactions on a square lattice, according to the mapping
\begin{equation}
    \label{eq:triangular_to_square}
    \diagram[0.9]{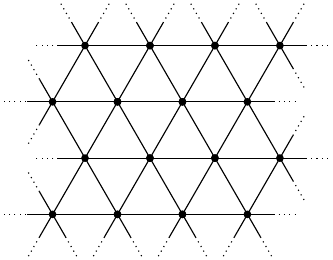}
    \quad \mapsto \quad
    \diagram[0.9]{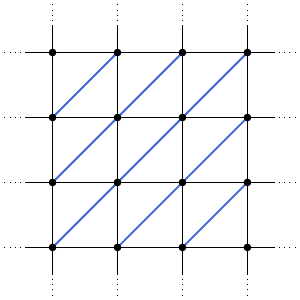}.
\end{equation}
Here, the NNN are represented by the diagonal blue lines.

Furthermore, we will work at a fixed particle density of $n = 1$, corresponding to half-filling.
To impose the finite particle density, we use an iPEPS ansatz with an internal $f\Z_2 \times \U{1}$ symmetry, corresponding to the product of the fermion parity and particle number symmetries of the Hamiltonian.
The chemical potential $\mu$ is set to zero.
To construct a state with a finite fermionic charge density, we should use local tensors with a non-trivial total physical charge corresponding to an odd fermion parity and $+1$ particle number.
This net physical charge is imposed by adding an auxiliary one-dimensional physical space with odd fermion parity and $-1$ particle number for every local PEPS tensor, which exactly compensates the offset in the physical charge.
In practice, this is equivalent to shifting the physical charges of the Hamiltonian itself by absorbing an identity on the auxiliary space using an appropriate isometry,
\begin{equation}
    \label{eq:peps_charge_density}
    \diagram[0.9]{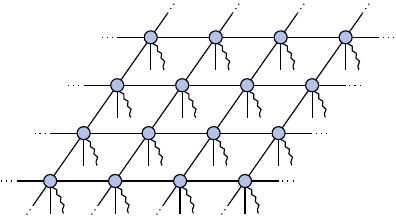},
    \qquad
    \diagram[1.0]{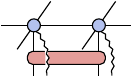} \cong \diagram[1.0]{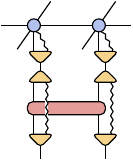}.
\end{equation}
By implementing the finite charge density as a charge shift of the Hamiltonian, we can directly use the contraction and optimization tools as outlined in \cref{sec:overview} without having to modify the ansatz or any of the corresponding contractions.
Importantly, because the auxiliary space is one-dimensional, this does not affect the performance at all.

To verify this approach, we optimize the ground state of the Fermi-Hubbard model on the triangular lattice at half-filling for different values of $U / t$, using an iPEPS defined on a $3 \times 3$ unit cell and a $f\Z_2 \times \U{1}$ symmetry.
To determine the charge distribution of the iPEPS virtual space, just as in \cref{sec:j1j2}, we use simple update with a suitable SVD truncation rank to dynamically select the appropriate $f\Z_2 \times \U{1}$ charges for a given bond dimension $D$.
We perturb the simple update result while retaining the charge structure to obtain a suitable initial iPEPS.
Given this initial iPEPS, we initialize a contraction environment by starting from a product state environment and running a few CTMRG iterations using an SVD with truncation rank $\chi > 3 D^2$.
As in~\cref{sec:j1j2}, this procedure dynamically determines the environment charge distributions.
During the L-BFGS optimization, the iPEPS virtual spaces are kept fixed.
The environment virtual spaces are kept fixed within a single energy evaluation in the optimization procedure.
Every few L-BFGS iterations, the charge distribution of the environment is updated by performing an intermediate CTMRG contraction similar to the environment initialization, where we use a fixed SVD truncation rank $\chi$ but allow the charge distribution to vary.
We run the optimization for 500 L-BFGS iterations, which is sufficient to converge the energy density to good accuracy.
After the optimization, we re-gauge the final iPEPS and compute a final CTMRG environment with a larger bond dimension of $1.5 \chi$ used to evaluate observables.
Just as before, we do not perform a full variational optimization, but rather break off the optimization at a gradient norm of order $\|\nabla E\| \sim 10^{-4}$.
This means the resulting final energies evaluated according to the procedure outlined here give an upper bound on the true variational energy, which is sufficient for the simple benchmarks presented here.

\begin{figure}
    \centering
    \includegraphics[width=0.95\linewidth]{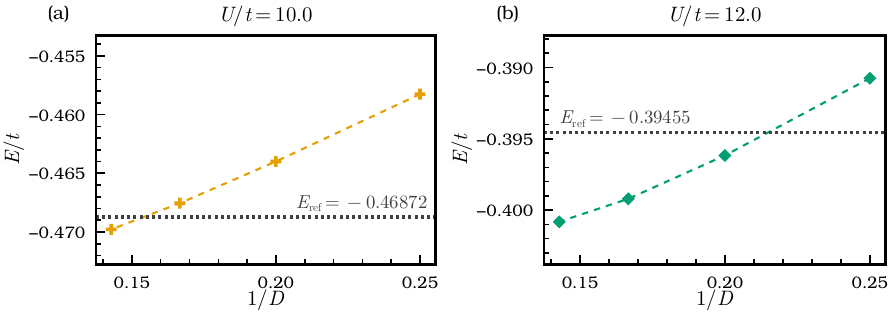}
    \caption{Optimized energies as a function of inverse iPEPS bond dimension (corresponding to $D = 4, 5, 6, 7$) for the Fermi-Hubbard model on the triangular lattice, for (a) $U / t = 10.0$ and (b) $U / t = 12.0$. Reference energies computed using a finite bond dimension extrapolation from DMRG results on finite cylinders of 72 sites are indicated as dotted lines \cite{chen_quantum_2022}.}
    \label{fig:triangular_energy}
\end{figure}

In \cref{fig:triangular_energy} we show the energy per site for intermediate values $U / t = 10.0, 12.0$ at various bond dimensions $D$, comparing to benchmark values computed with DMRG \cite{chen_quantum_2022}.
The DMRG reference results correspond to energies on finite cylinders of 72 sites, extrapolated in bond dimension.
Since our results and the reference energies are still sensitive to finite bond dimension and finite size effects respectively, we note that \cref{fig:triangular_energy} should only be interpreted as a qualitative confirmation that we obtain similar results.

\begin{figure}
    \centering
    \includegraphics[width=0.95\linewidth]{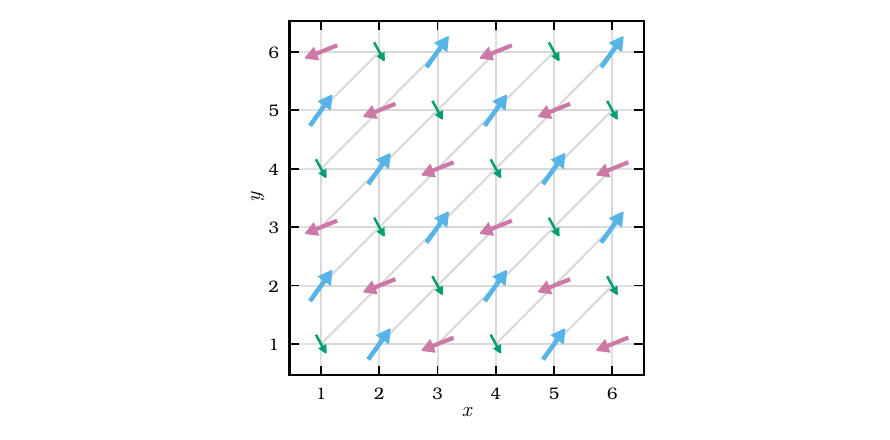}
    \caption{Planar projection of the magnetization in the Fermi-Hubbard model on the triangular lattice at $U/t = 30$, as a function of lattice coordinates.}
    \label{fig:triangular_magenetization}
\end{figure}

In addition, we performed a full variational optimization with $\|\nabla E\| < 10^{-7}$ at $D = 4$ closer to the Heisenberg limit at $U/t = 30$.
\cref{fig:triangular_magenetization} shows the planar projection of the calculated magnetization for the optimized ground state, which exhibits the frustrated $120^\circ$ N\'eel order as expected.

\subsection{Benchmarking (non-)Abelian symmetries}
\label{sec:benchmarks}

Our final example is a performance benchmark of iPEPS algorithms in the presence of different internal symmetries.
Concretely, we consider the variational optimization of the iPEPS ground state of the $\SU{3}$ Fermi-Hubbard model on the square lattice.
Its Hamiltonian still has the form of \cref{eq:fermi-hubbard}, except that $\sigma$ now sums over three $\mathrm{SU}(3)$ flavors $\{1, 2, 3\}$ \cite{bird_partial_2025,kleijweg_zigzag_2025}.

To set up the benchmarks, we first optimize the iPEPS ground state of the $\SU{3}$ Fermi-Hubbard model at $U / t = 4.0$ for a range of bond dimensions.
Since this point lies deep in the metallic phase, we use an iPEPS ansatz with a $3 \times 3$ unit cell which has the full non-Abelian $f\Z_2 \times \U{1} \times \SU{3}$ symmetry of the model.
We also impose a finite particle density of $n = 1$ fermion per site using the procedure of \cref{eq:peps_charge_density}, where we add a one-dimensional auxiliary space with odd fermion parity, $-1$ particle number and singlet $\SU{3}$ charge.
We use precisely the same procedure as in \cref{sec:triangular_hubbard} for determining the charge distribution of the initial iPEPS and environment virtual spaces, and periodically updating the virtual space structure during the optimization.
Here, we choose the environment bond dimension as $\chi = D^2$.
For each iPEPS bond dimension $D$, we generate a trial state which is representative of the ground state by optimizing for 100 L-BFGS iterations.

\begin{figure}
    \centering
    \includegraphics[width=0.95\linewidth]{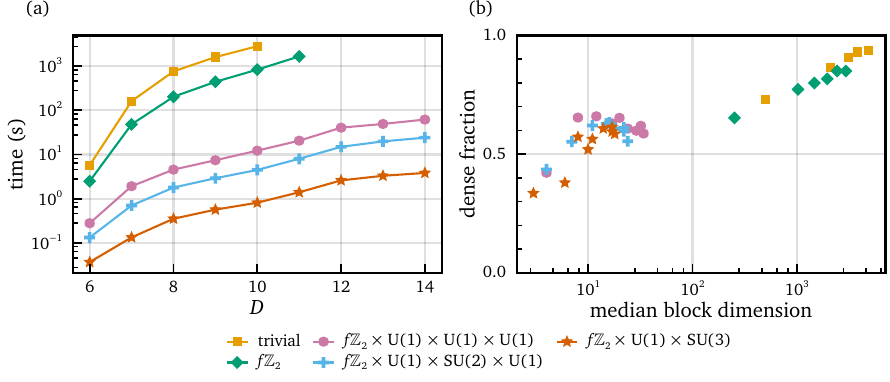}
    \caption{(a) Time (in seconds) per CTMRG iteration for contracting the ground state of the $\SU{3}$ Fermi-Hubbard model at $U / t = 4.0$, as a function of iPEPS bond dimension $D$ for different internal symmetries. The environment bond dimension is chosen as $\chi = D^2$. (b) Fraction of the total time per iteration spent on dense (BLAS/LAPACK) matrix operations, as a function of the median block size of a symmetric enlarged corner tensor as it appears on the  left hand side of \cref{eq:ctmrg}, for different internal symmetries.}
    \label{fig:su3_benchmark}
\end{figure}

Given the charge distribution of the optimized trial state and its environment, we then record the wall time in seconds required for a single CTMRG iteration\footnote{The specific CTMRG algorithm we chose here computes a full SVD of an environment made up of two enlarged corners on the left-hand side of \cref{eq:ctmrg}. Different flavors of CTMRG give exactly the same qualitative behavior of the cost scaling for different symmetries.} using random initializations with the same iPEPS and environment virtual spaces.
We then repeat this procedure for a range of reduced symmetries.
In particular, starting from the $f\Z_2 \times \U{1} \times \SU{3}$-symmetric spaces of the optimized trial state, we convert these to equivalent spaces with a reduced symmetry, and benchmark a single CTMRG iteration using a random iPEPS and environment of the correct shape.
We do this for a sequence of four different internal symmetries, starting from the full $f\Z_2 \times \U{1} \times \SU{3}$ symmetry, then reducing to a $f\Z_2 \times \U{1} \times \SU{2} \times \U{1}$, $f\Z_2 \times \U{1} \times \U{1} \times \U{1}$, $f\Z_2$, and finally a trivial symmetry.
This latter non-symmetric point is added as a reference only, as breaking the fermionic symmetry in a physical system is not possible.
The benchmarks were performed using 8 threads on an AMD EPYC 7552 processor, and are shown in \cref{fig:su3_benchmark}.
We clearly see from \cref{fig:su3_benchmark}(a) that imposing the full non-Abelian symmetry can give a speedup of an order of magnitude compared to just imposing Abelian symmetries.
In our benchmark setup, the only place where multithreading was used is in the underlying BLAS and LAPACK routines for matrix multiplication and linear algebra.
For the bond dimensions shown in \cref{fig:su3_benchmark}(a), we note that the matrix blocks involved in the operations on symmetric tensors for the larger symmetry groups $f\Z_2 \times \U{1} \times \SU{3}$,  $f\Z_2 \times \U{1} \times \SU{2} \times \U{1}$ and $f\Z_2 \times \U{1} \times \U{1} \times \U{1}$ were too small to activate this multithreading, meaning only the $f\Z_2$ and trivial symmetries actually benefited from using multiple threads.
Therefore, at the bond dimensions used here, the speedup due to imposing non-Abelian symmetries would be significantly larger when running on a single thread.

We do however mention that pushing the non-Abelian performance even further requires some more technical improvements, which is still a work in progress.
As the symmetry group becomes larger, so does the overhead of the associated bookkeeping.
Nevertheless, this information depends only on the structure of the vector spaces, and not on the entries, and can therefore be cached.
This gives great performance for repeated CTMRG iterations with fixed bond dimensions, as then the bookkeeping cost is amortized over the different iterations, as shown in the results of \cref{fig:su3_benchmark}.
However, naively running a CTMRG procedure with variable bond dimensions can become prohibitively expensive.
Since the caching becomes less efficient, either the memory requirements grow too much or too much of the bookkeeping has to be recomputed, leading to longer run times.

\begin{figure}
    \centering
    \includegraphics[width=0.95\linewidth]{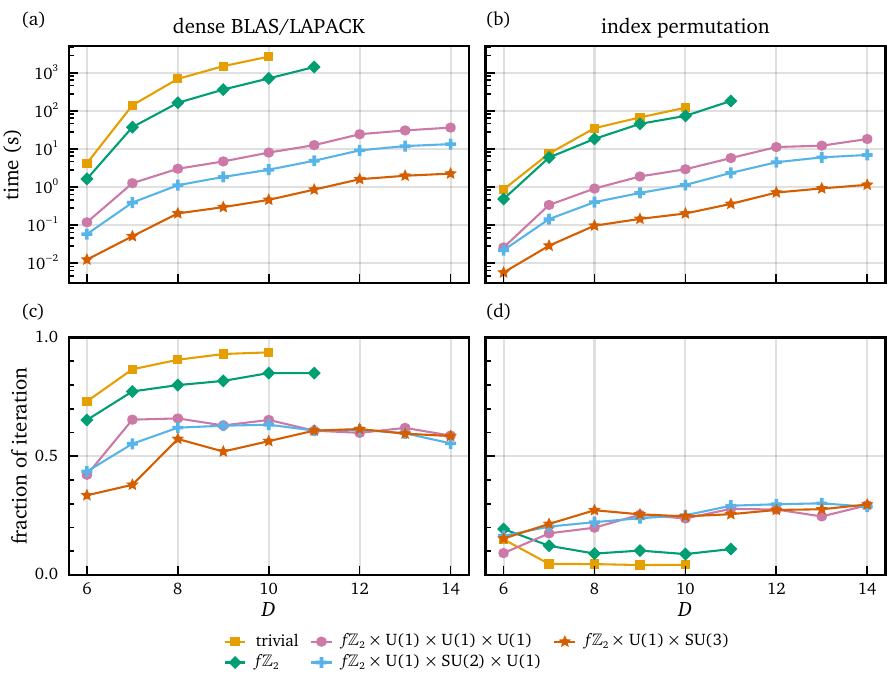}
    \caption{Time per CTMRG iteration spent (a) on dense BLAS and LAPACK operations on the blocks of the symmetric tensors and (b) on index permutations, including the fusion-tree recoupling for non-Abelian symmetries, for the same setup as \cref{fig:su3_benchmark}. (c), (d) The same quantities as a fraction of the total time per iteration.}
    \label{fig:su3_split}
\end{figure}

To investigate this further, we split the time of a single CTMRG iteration into the time spent on dense linear algebra, and the time spent on permuting tensor indices.
The dense linear algebra contribution is comprised of the BLAS matrix multiplications and LAPACK singular value decompositions acting on the individual blocks of the symmetric tensors.
For the contribution of permuting tensor indices, we count both the time spent on moving data between blocks, as well as the time spent on applying the fusion-tree recoupling coefficients of non-Abelian symmetries.
For now, we exclude the time spent on associated symmetry bookkeeping that can be cached and reused, first focusing on operations that are performed on the actual tensor entries during hot loops of tensor network algorithms.
These split timings are shown in \cref{fig:su3_split}.
We see that all symmetries are dominated by dense linear algebra, while the larger symmetry groups show more significant contributions from the index permutations.
That effect also does not vanish for the bond dimensions that we probe in our benchmarks, which is best explained by \cref{fig:su3_benchmark}(b) that shows that the median block dimension remains orders of magnitude smaller compared to the smaller symmetry groups.

\begin{figure}
    \centering
    \includegraphics[width=0.95\linewidth]{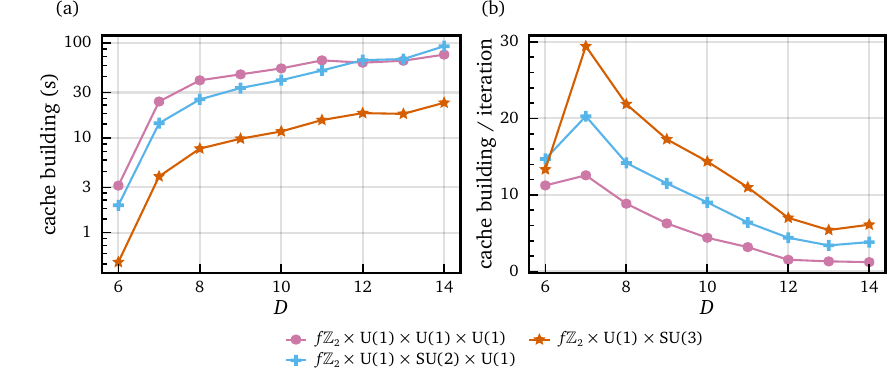}
    \caption{(a) The total time required to compute all the symmetry structure data, that is cached and reused during CTMRG iterations, from scratch, as a function of the PEPS bond dimension for different symmetries. (b) The same cache building time, expressed relative to the time of a single CTMRG iteration for the same bond dimension and symmetry.}
    \label{fig:su3_cache}
\end{figure}

As the final part of the benchmark, we also report the time spent on symmetry bookkeeping that was cached and reused during the CTMRG iterations benchmarked above.
To measure this effect, we provide worst-case-scenario timings, starting from a fully cold cache for every iteration, while in reality some of the cached information can be reused even for different tensor sizes.
\cref{fig:su3_cache} shows the total time required to generate the full set of symmetry structure data used during a CTMRG iteration, both in absolute time and relative to the corresponding iteration time.
These timings were recorded using 8 Julia threads, since cache generation can be accelerated significantly through parallelization.
While this relative cost can be substantial at small bond dimensions, it quickly settles into a controllable overhead as bond dimensions are increased.
In particular, in the context of a full variational optimization where a PEPS with the same spaces is being contracted repeatedly, this small upfront cost is far outweighed by the time it saves over the whole optimization.
The memory size of the symmetry data cache itself is, however, non-negligible, making it an additional aspect that has to be considered when managing the memory pressure associated with large-scale simulations using symmetric PEPS.
With increasing bond dimensions, the memory size of the symmetry data cache tends to be of the same order as the peak memory required for the live tensor data during an algorithm.

%% file: 4-conclusion.tex
\section{Conclusion}
\label{sec:conclusion}

We have presented PEPSKit.jl, an open-source Julia package for simulating 2D quantum many-body systems using infinite projected entangled-pair states.
The package provides a unified framework covering the three main components of most state-of-the-art PEPS workflows: approximate contraction of infinite tensor networks via CTMRG and boundary-MPS methods, time evolution through Trotter decomposition, and variational ground-state optimization via automatic differentiation.
A central design goal has been to make these capabilities broadly accessible without sacrificing performance: by building on TensorKit.jl and the wider QuantumKitHub ecosystem, PEPSKit.jl inherits full support for Abelian, non-Abelian and fermionic symmetries, all within a composable and extensible software framework.

The examples presented in Section \ref{sec:examples} illustrate the practical reach of the package across a range of physically relevant settings, including finite-temperature spin models, frustrated magnets, and fermionic systems on non-trivial lattice geometries.
This demonstrates how combining AD-based optimization with symmetry enforcement yields both competitive numerical accuracy and significant computational speedups, and that these gains are accessible without bespoke, model-specific implementations.

We believe that open-source research software plays an indispensable role in the future development of the tensor network community and the wider quantum many-body physics field.
The relative maturity of the MPS ecosystem has substantially lowered the barrier to entry for 1D problems, enabling rapid exploration of new models, algorithms, and physical phenomena.
PEPSKit.jl aims to help close this gap for 2D systems, providing both a production-ready set of algorithms for practitioners and a flexible platform on which new algorithmic ideas can be prototyped and tested.